\documentclass[pre,preprint,showpacs]{revtex4}
\usepackage[dvips]{graphicx}
\usepackage{latexsym,amsmath,amssymb,amsthm,xspace,times}
\usepackage{color}
\begin{document}
\title{Effects of jamming on non-equilibrium transport times in nano-channels}

\author{A. Zilman$^{1,2}$, J. Pearson$^{1}$ and G. Bel$^{2,3}$}
\affiliation{
$^{1}$Theoretical Biology and Biophysics Group and $^{2}$Center for Nonlinear Studies,
Theoretical Division\\
$^{3}$Computer, Computational and Statistical Sciences Division\\
Los Alamos National Laboratory, Los Alamos, NM 87545, USA }
% Revision date - uncomment to exclude date in the final version

\begin{abstract}
Many biological channels perform highly selective transport without direct input of metabolic energy and without transitions from a 'closed' to an 'open' state during transport. Mechanisms of selectivity of such channels serve as an inspiration for creation of artificial nano-molecular sorting devices and bio-sensors. To elucidate the transport mechanisms, it is important to understand the transport on the single molecule level in the experimentally relevant regime when multiple particles are crowded in the channel. In this paper we analyze the effects of inter-particle crowding on the non-equilibrium transport times through a finite-length channel by means of analytical theory and computer simulations.
\end{abstract}
\maketitle
\section{Introduction}
The functioning of living cells depends critically on molecular transport through various transport channels \cite{stein-book}. Many of them function without a direct input of metabolic energy and without a movable 'gate' that would involve transitions from an 'open' to a 'closed' state during transport. Nevertheless, such channels are selective, efficient and fast. Examples include porins, Nuclear Pore Complex and others \cite{aquaporins-book,bezrukov-kullman-maltoporin-2000,mike_review,wente-review,schulten_glycerol}. The functioning of such channels has served as an inspiration for the creation of artificial biosensors and nano-molecular filters \cite{martin-DNA-2004,polymer-nanotubes-2008,martin-apoenzymes-1997,caspi-elbaum-2008,tijana,akin-DNA-nanotubes-2007,martin-antibody-science-2002} that promise to play an ever increasing role in nano-technological and nano-medical applications, such as single-mismatch DNA detection \cite{martin-DNA-2004,akin-DNA-nanotubes-2007},  enantiomer separation \cite{martin-antibody-science-2002}, pathogen detection \cite{bezrukov-anthrax} and design of  antibiotic drugs optimized for penetrating the cell \cite{bezrukov-antibiotics-PNAS-2002}. Such man-made channels  also serve as testbeds for examining models of biological transport \cite{tijana,caspi-elbaum-2008}.

Biological and artificial transport channels, such as those mentioned above, usually contain a passageway through which the molecules translocate by diffusion. From recent experimental and theoretical work, it has become increasingly clear that in many cases the transport selectivity of such channels is not dictated merely by molecule size, but is controlled by transient binding of the transported molecules inside the passageway \cite{bezrukov-kullman-maltoporin-2000,schulten_glycerol,bezrukov-antibiotics-PNAS-2002,bezrukov-asymmetric-2007,noble-theory-1991,aquaporins-book,zilman-BJ,mike_review,wente-review}. The crucial insight into understanding the transport selectivity of such channels is that even in the absence of any physical barrier for the entrance to the channel, the probability of a particle to  translocate through it is low (of an order of the aspect ratio of the channel) \cite{berezhkovskii_ptr}. Transient trapping (due to binding) inside the channel overcomes this 'dimensionality barrier' \cite{berezhkovskii_ptr,we-NPC-plos-2007,bezrukov-asymmetric-2007,bezrukov-sites-2005}. However, if the molecules are trapped in the channel for too long, the channel becomes crowded and transport is diminished. The interplay of these two effects provides a basis for selective transport, whereby only the molecules that are trapped in the channel for an optimal time transit through the channel with a high flux \cite{bezrukov-sites-2005,berezhkovskii-optimal-2005,we-NPC-plos-2007,strange-people-occusion-pnas-2006}. Related mechanisms have been known in the context of carrier-assisted membrane transport as 'facilitated diffusion' \cite{myoglobin-1966,noble-theory-1991,cussler-book-1}. Theoretical models that include the transient trapping combined with the effects of confinement \cite{bezrukov-sites-2005,berezhkovskii-optimal-2005,we-NPC-plos-2007,strange-people-occusion-pnas-2006,noble-theory-1991,AK-SK-2008} provide a good explanation of the behavior of the mean flux through nano-channels and show a good agreement with the experimental data \cite{zilman-BJ}.

However, from a  biological perspective, transport of a single molecule can constitute a significant signalling effect \cite{wente-review,mike_review}. Thus, it is important to understand the transport through such  channels on the single molecule level. Advances in fluorescent microscopy and other methods allow one to follow the transport of individual molecules through a channel \cite{yang-musser-JCB-2006,kubitschek_single,werner3d,kubitscheck-2008,akin-DNA-nanotubes-2007,musser_single}. Single molecule tracking experiments provide a wealth of information about the transport mechanisms, which is not accessible from the measurements of the bulk flux through the channel. The kinetics of transport of a single particle through the channel in the absence of other particles is well understood \cite{redner-book,gardiner-book,berezhkovskii-times,berezhkovskii_ptr}. In this paper, we analyze the effects of crowding of the particles inside the channel on the transport times of individual particles in the experimentally relevant regime when a non-equilibrium steady state flux passes through the channel.
% The transport of individual particles takes place on the background of non-equilibrium steady state flux through a channel. The transported particles interact and interfere with each other's passage.   Effect of jamming on the diffusion properties has been studies before in mean field (effective medium) and beyond and by simulations.
\section{Single Particle}
%\textit{Single particle:}
Here, we briefly review the kinetics of a single particle passing through the channel in order to explain the methods employed herein. The channel is represented as a sequence of 'sites' $1,...,N$. Inside, the particle  performs diffusion-like random walk starting at the 'entrance' site $1$ and hopping between the internal sites $1\leq i \leq N$ at an average rate $r$ (for simplicity, we assume that the channel is uniform). The particle can leave the channel from the terminal sites $1$ and $N$ with an average rate $r_o$.   Transient trapping in the channel is described by choosing $r_o<r$. This hopping process is illustrated in Fig. 1. %\ref{fig-schematic-prl}.
\begin{figure}[htbp]\label{fig-schematic-prl}
\centerline{
\includegraphics[width=\linewidth]{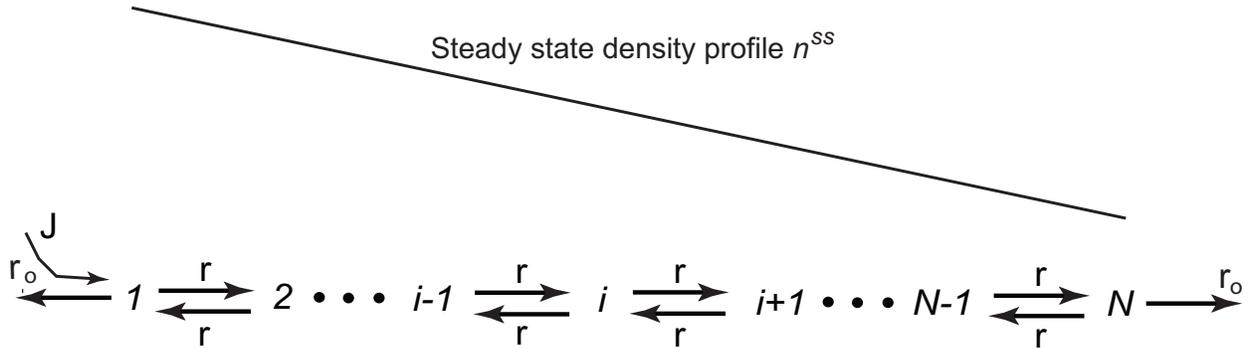}
}
\caption{The channel is represented by a sequence of sites $1,...,N$ between which the particles can hop with rate $r$. The rate of hopping out of the channel from its ends is $r_o$. In the single particle case, a particle starts at site $1$ and hops inside the channel until it exits from either end. In the multi-particle case, the particles enter at site $1$ with an average rate $J$, if its occupancy is less than the maximal allowed. The line shows the steady state concentration profile.}
\end{figure}
%%%%%%
At any time $t$, the position of the  particle in the channel is described by the vector of probabilities $p_i(t)$ to be at a particular site $i$:
$|p(t)\rangle=(p_1(t),...p_i(t)...p_N(t))$. We also define the vector $|i\rangle$ as a vector with the $i'th$ element equal to $1$ and all other elements equal to $0$, so that $\langle i|p(t)\rangle= p_i(t)$ (where $\langle x\vert y \rangle$ is the scalar product of the vectors $\vert x\rangle$ and $\vert y\rangle$).
%%%%%%%
The Master equation for the probability vector, describing the hopping through and out of the channel ends, can be written as (see the Appendix for details)
\begin{eqnarray}\label{equation-for-p-1-prl}
\frac{d}{dt}|p(t)\rangle=\hat{M}\cdot |p(t)\rangle.
\end{eqnarray}
The formal solution of the equation (\ref{equation-for-p-1-prl}) can be written as $|p(t)\rangle=e^{\hat{M}t}|p(0)\rangle $, where $|p(0)\rangle$ is the initial condition \cite{gardiner-book,pearson-BJ-2005}; for a particle starting at site $1$, $|p(0)\rangle=\vert 1 \rangle=(1,0...0)$.
The instantaneous probability flux to the right out of the channel is $r_op_N(t)$, and the probability that the particle had exited the channel from the right side by time $t$ is  $P_{\rightarrow}^t=\int_0^t r_op_N(t')dt'$ \cite{berezhkovskii_ptr,gardiner-book}.

The total probability to exit to the right $P_{\rightarrow}\equiv P_{\rightarrow}^{\infty}$ is
%%%%
\begin{equation}\label{equation-Pr-single-prl}
P_{\rightarrow}=\int_0^{\infty}r_o\langle N|e^{\hat{M}t}|1\rangle dt'=-r_o\langle N|\hat{M}^{-1}|1\rangle.
\end{equation}
%%%%
where $\langle i|\hat{X}|j\rangle\equiv \hat{X}_{ij}$. After some algebra (see the Appendix for details), equation (\ref{equation-Pr-single-prl}) gives for the total probability to exit from the channel on the right (the translocation probability): $P_{\rightarrow}=\frac{1}{2+(N-1)r_o/r}$, in accord with previous works \cite{berezhkovskii_ptr,zilman-BJ}.
Note that $P_{\rightarrow}$ increases as $r_o$ diminishes. That is, trapping of the particle in the channel increases the translocation probability \cite{myoglobin-1966,cussler-book-1,noble-theory-1991,we-NPC-plos-2007,berezhkovskii_ptr}.

We now calculate the directional mean exit times. The probability distribution of the exit times to the right $f_{\rightarrow}(t)$ is
$f_{\rightarrow}(t)=-\frac{1}{P_{\rightarrow}}\frac{d}{dt}(1-P_{\rightarrow}^t)=r_op_N(t)/P_{\rightarrow}$ \cite{berezhkovskii-times,gardiner-book,redner-book}. Thus, the mean time to exit the right is
\begin{equation}
\overline{T}_{\rightarrow}=\int_0^{\infty}t' f_{\rightarrow}(t')dt'=r_o\langle N|\left(\hat{M}^{-1}\right)^{2}|1\rangle/P_{\rightarrow}.
\end{equation}
Similarly, the mean first passage time to the left is
\begin{equation}
\overline{T}_{\leftarrow}=\int_0^{\infty}t' f_{\leftarrow}(t')=r_o\langle 1|\left(\hat{M}^{-1}\right)^{2}|1\rangle/P_{\leftarrow}.
\end{equation}
The mean time to exit from \textit{any} of the ends is
\begin{equation}\label{AT}
\overline{T}=r_o\int_0^{\infty}t\left(p_N(t)+p_1(t)\right)dt=\overline{T}_{\leftarrow}P_{\leftarrow}+\overline{T}_{\rightarrow}P_{\rightarrow}.
\end{equation}
Using the equations above, we obtain explicit expressions for the mean times:
\begin{align}
\overline{T}_{\rightarrow}&=\frac{N \left(6P_{\leftarrow}+P_{\rightarrow}\left(N(N-3)+2\right)(r_o/r)^2\right)}{6r_o},\nonumber\\
\overline{T}_{\leftarrow}&=\frac{N \left(6P_{\leftarrow}P_{\rightarrow}+P^{2}_{\rightarrow}\left(N(2N-3)+1\right)(r_o/r)^2\right)}{6r_oP_{\leftarrow}},\nonumber\\
\overline{T}&=\frac{N}{2r_o},
\end{align}
in agreement with previous results obtained in the continuum limit \cite{berezhkovskii-times,bezrukov-sites-2005}.
Note that the mean trapping time $\overline{T}$ is \emph{linearly} proportional to the channel length $N$. Surprisingly, the mean time for the particle to exit to the left $\overline{T}_{\leftarrow}$ also scales like $N$ for $N\gg 1$, due to the possibility of large excursions into the channel before it returns to the left end. By contrast, the mean exit time  to the right has two distinct regimes. For short channels, or strong trapping ($Nr_o/r\ll 1$), $\overline{T}_{\rightarrow}\sim \frac{N}{2r_o}$, while for long channels, or weak trapping, ($Nr_o/r\gg 1$), $\overline{T}_{\rightarrow}\sim \frac{N^2}{2r}$ (see also Fig. 3). Physically, for strong trapping, the bottleneck for the exit to the right is the release from the channel end, while for long channels and weak trapping the exit time is dominated by the time it takes to diffuse through the channel from left to right.
\section{Single particle on the background of the steady state flux}
%\textit{Single particle on the background of the steady state flux}:
When a finite flux $J$ impinges onto the channel entrance,  at any moment there can be many particles in the channel that might interfere with each other's passage and prevent the entrance of new ones. The particles in the channel obey the same kinetics as the single particles, with a condition that a site can contain up to a maximal number of particles $m$. Following \cite{zilman-BJ,chou-PRL-single-file-1998,rittenberg,schutz-review-2005,macdonald-gibbs-1968}, the system can be described in terms  in terms of site occupancies $n_i=|n\rangle^{ss}_i $. For constant $J$, a non-equilibrium steady state is established. The steady state profile of a uniform channel can be solved exactly: $n^{ss}_i=\frac{JP_{\rightarrow}(1+(N-i)r_o/r)}{r_o+JP_{\leftarrow}/m}$ \cite{zilman-BJ,chou-PRL-single-file-1998,kutner-tracer,rittenberg,schutz-review-2005}.

We now turn to the main results of this paper - how does the crowding, when many particles are present in the channel, affect  the transport times of individual particles within the non-equilibrium steady-state flux. To the best of our knowledge, no exact analytical solution exists in this case. The transport of an individual particle can be viewed as occurring on the background of the steady state density profile $|n\rangle^{ss}$. In the mean field  approximation, the probability $p_i(t)$ of a particle to be present at a given site is described by the following equations \cite{granek-nitzan,chou-tasep-2003,macdonald-gibbs-1968,kutner-tracer}:
\begin{eqnarray}
\frac{dp_{i}}{dt}&=&rp_{i-1}(1-\frac{n^{ss}_i}{m})+rp_{i+1}(1-\frac{n^{ss}_i}{m}) \label{kinetics_with_exclusion-ss-prl}\\
&&-rp_i(1-\frac{n^{ss}_{i-1}}{m})-rp_i(1-\frac{n^{ss}_{i+1}}{m}),\nonumber
\end{eqnarray}
with the appropriate boundary conditions (see the Appendix for details). Using matrix notations:
\begin{eqnarray}
\frac{d}{dt}|p(t)\rangle=\hat{M}^{ss}\cdot |p(t)\rangle. \label{equation-for-n-1-ss-prl}
\end{eqnarray}
Explicit matrix elements of $\hat{M}^{ss}$ are given in the Appendix. As in the single-particle case above, the linear equations (\ref{kinetics_with_exclusion-ss-prl},\ref{equation-for-n-1-ss-prl}) can be solved analytically.

To test the feasibility of the mean field approximation, we compared the probability of a particle to exit to the right, computed  using the exact solution for the steady state density  with the mean field result (see below). First, the average exit flux to the right is  $J_{\rightarrow}=r_on^{ss}_N$, which yields for the probability of an individual particle within this steady state flux to exit to the right \cite{zilman-BJ}:
\begin{equation}\label{p-exit-ss-prl}
P^{ss}_{\rightarrow}=\frac{J_{\rightarrow}}{J(1-n_1/m)}=\frac{1}{2+(N-1)r_o/r}.
\end{equation}
On the other hand, from the mean field approximation of  eq.~( \ref{kinetics_with_exclusion-ss-prl}) $P^{ss}_{\rightarrow}=-r_o\langle N|(\hat{M}^{ss})^{-1}|1\rangle$ (see Eq.~(\ref{equation-Pr-single-prl})). Using the expressions for $n_i^{ss}$, after some algebra we get the same result as the exact expression, eq.(\ref{p-exit-ss-prl}). Thus, the mean field approximation yields an exact result: the probability of an individual particle to exit to the right is not affected by crowding and is the same as in the single-particle case (at least for uniform channels) \cite{kutner-tracer,schutz-review-2005}. The directional mean exit times can be calculated by repeating the same algebra as for the case of a single particle, but with $M^{ss}$ instead of $M$. We find that the mean trapping time is $\overline{T}^{ss}=\frac{N}{2r_o} $
- surprisingly, like the translocation probability, the mean trapping time is also not affected by the crowding.
By contrast, the directional times to exit to the right and to the left, $\overline{T}^{ss}_{\rightarrow},\overline{T}^{ss}_{\leftarrow}$
respectively, do change due to inter-particle interactions, compared to the single particle case. After some algebra (see the Appendix for details), one gets for the mean time to exit to the left:
\begin{widetext}
\begin{align}
& \overline{T}^{ss}_{\leftarrow}=r_o\langle 1| (M^{ss})^{-2}|1\rangle/P_{\leftarrow} \\ &=\frac{rm\left(JP_{\leftarrow}+mr_o\right)^2}{ J^3 r_oP_{\leftarrow}P_{\rightarrow}}  \left[\psi\left(N+\frac{rm}{JP_{\rightarrow}}\right)-\psi\left(\frac{rm}{JP_{\rightarrow}}\right)\right]-\frac{Nm}{ J^2 P_{\leftarrow}} \left[m r_o+JP_{\leftarrow}+\frac{J}{2}\right],\nonumber
\end{align}
\end{widetext}
where $\psi(x)=\frac{d}{dx}\Gamma(x)$; $\Gamma(x)$ is a $\gamma$-function. The mean exit time to the right can be obtained in a similar fashion, using equation~(\ref{AT}).
The dependence of the exit times on the impinging flux $J$ is illustrated in Fig. 2. Unlike the exit probabilities, crowding increases the mean time to exit to the right $\overline{T}_{\rightarrow}$, and decreases the mean time to exit to the left $\overline{T}_{\leftarrow}$. Interestingly, however, the qualitative dependence on the channel length $N$ is similar to the single-particle case (J=0), as shown in Fig. 3. Importantly, the transport times remain finite  even in the fully jammed regime ($J\rightarrow \infty$), when the flux through the channel saturates to its maximal value. In particular, the mean time to exit to the right tends to $
\frac{N-2P_{\leftarrow}^2}{2P_{\rightarrow}r_o}$, while the mean time to exit to the left tends to $\frac{P_{\leftarrow}}{r_o}$ (see the Appendix for details).
\begin{figure}[htbp]\label{fig-mean-times-J}
\centerline{
\includegraphics[width=\linewidth]{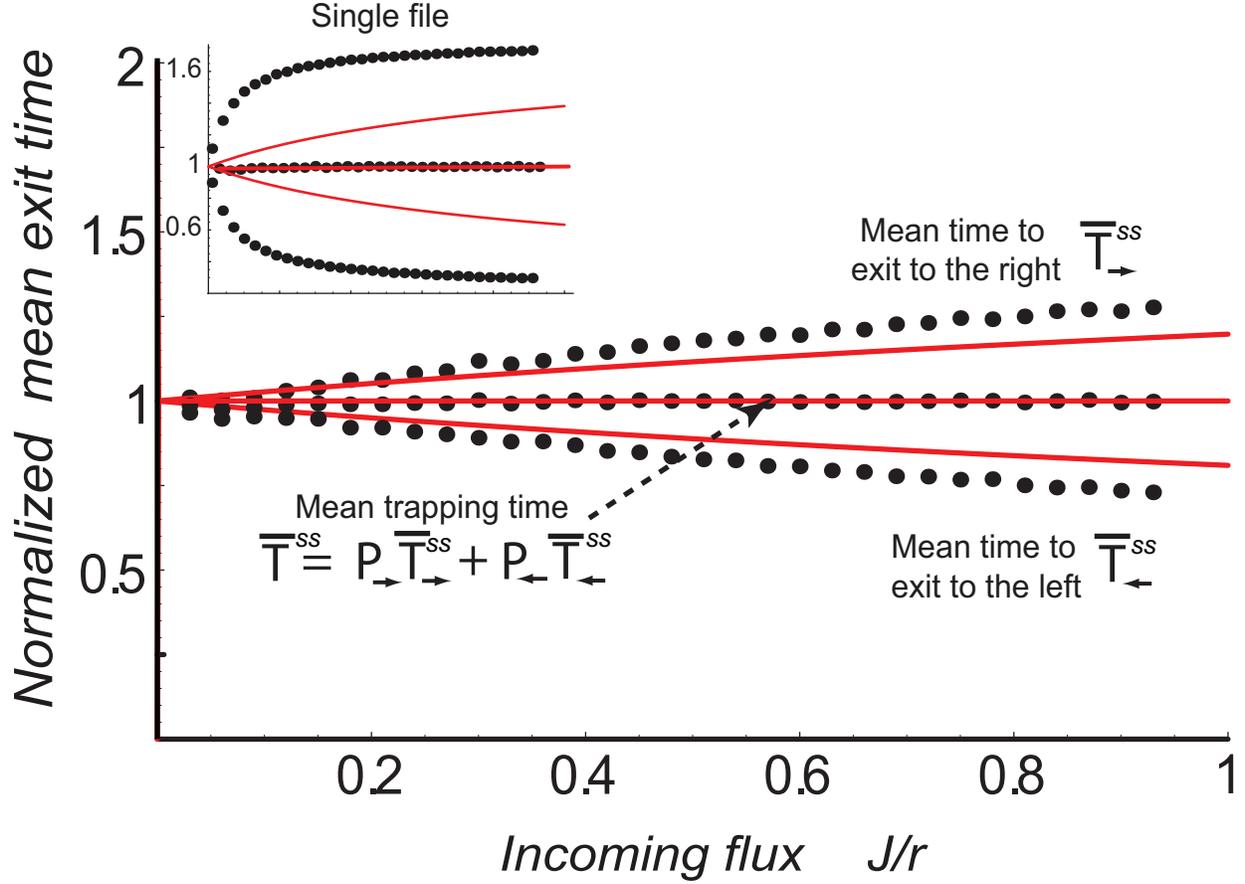}
}
\caption{ Ratios of the mean exit times in the jammed regime to the single-particle times. Upper lines - $\overline{T}^{ss}_{\rightarrow}/\overline{T}_{\rightarrow}$. Lower lines - $\overline{T}^{ss}_{\leftarrow}/\overline{T}_{\leftarrow}$. Solid red lines - analytical solution, dotted lines - simulations; $N=6$, $r=1$, $r_o=0.1$, $m=3$ for all lines.
\emph{Inset:} Same for single file transport $m=1$. }
\end{figure}

\begin{figure}[htbp]\label{fig-mean-times-N}
\centerline{
\includegraphics[width=\linewidth,height=0.6\linewidth]{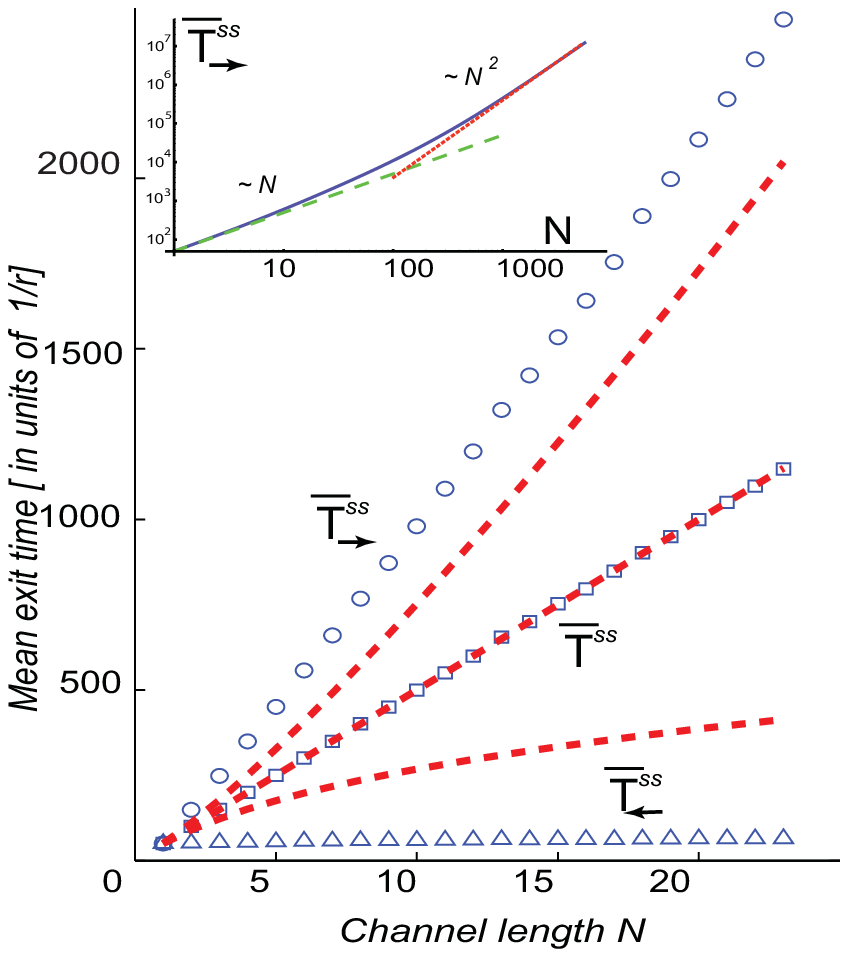}
}
\caption{Dependence of the transport time in the crowded regime on the channel length $N$. Dashed red lines - mean field results; symbols - corresponding simulations; $r=1$, $r_o=0.01 $, $J=0.5$. The dependence on $N$ is qualitatively similar to the single particle case - see text. The inset shows the transition between small $N$ (linear dependence) and large $N$ (quadratic dependence) regimes; logarithmic scale; $r=1$, $r_o=0.01 $, $J=0.1$. }
\end{figure}

In order to corroborate the results of the mean field approximation and to investigate the limits of its validity, we performed computer simulations of the transport through the channel using a variant of the Kinetic Monte Carlo algorithm \cite{voter}. Both the simulations and the mean field results show that the forward times are increased due to the jamming while the backward times are decreased. The increase in forward exit times is easily understood considering the reduction in hopping rates inside the channel due to crowding. The origin of the decrease in the backward exit time is more subtle: the crowding increases the number of particles which hop backwards out of the channel immediately after their entrance. For wide channels that can accommodate more than one particle at each site, the mean field results for the directional transport times $\overline{T}_{\rightarrow}$ and $\overline{T}_{\leftarrow}$ agree closely with the simulations. For strictly single file channels ($m=1$), the mean field approximation underestimates the  actual value of the  exit time to the right  $\overline{T}_{\rightarrow}$ and overestimates the exit time to the left $\overline{T}_{\leftarrow}$, but still reproduces the right qualitative dependence of the times on the flux $J$ and other parameters. The reason for the underestimation of the time to exit to the right is  that  the mean field approximation neglects the correlation between successive jumps (a particle hopping to one of its neighbor sites leaves behind it a vacancy and thus has a higher probability to hop back to the same site in the next jump). Interestingly, the simulations show that the mean field result for the mean trapping time $\overline{T}$ is exact (at least for a uniform channel). The mean field approximation can be improved by taking into account the correlations in the jumping rates of the neighboring particles and the fluctuations of the density around its mean value. Correction to the mean field diffusion rate of a tracer particle in equilibrium conditions were calculated in \cite{granek-nitzan,karger} in the framework of effective medium theory. In general, such corrections to the mean field improve the approximation including our case (data not shown); however systematic analysis of such corrections lies outside the scope of the present work.
\section{Summary}
%\newline\emph{Discussion}:  
To summarize, we have analyzed the effects of crowding and inter-particle competition for space on the transport times through narrow channels of finite length under a non-equilibrium steady state condition. The results of the mean-field analysis are corroborated by computer simulations. We have shown that in uniform channels the jamming increases the forward exit time, while decreasing the backward exit time. Surprisingly, jamming does not affect the mean dwelling time in uniform channels. The situation might be different in non-uniform channels; however, the mean field approximation should  provide a qualitatively correct picture even in this case \cite{macdonald-gibbs-1968}, full discussion of which lies beyond the scope of the present work and will be discussed elsewhere. The model provides a theoretical framework for analysis of single molecule transport through biological and artificial nano-channels. The parameters of the model, the rates $r_o$ and $r$, can be related to the experimentally controlled factors such as diffusion coefficients inside and outside the channel and the binding affinity of the molecule in the channel.  It is also important to emphasize the difference between the results of this paper and the well studied case of tracer diffusion in infinite single-file channels \cite{barkai-2002}. Finally, we note that the methods of this work can be extended to treat arbitrary molecular signalling pathways, such as multi-step enzymatic reactions \cite{nemenman-sinitsyn,english-xie}, conformational transitions of ion channels \cite{pearson-BJ-2005} and other systems \cite{Chou}.
\newline The authors are thankful to R. Groger, I. Nemenman, B. Munsky, A. Perelson, K. Rasmussen, N. Sinitsyn for stimulating discussions. This research was performed under the auspices of the U.S. Department of Energy.
\appendix*
\section{Detailes of the calculations in the main text}
\subsection{Single particle}
At any time $t$, the position of the  particle in the channel is described by the vector of probabilities $p_i(t)$ to be at a particular site $i$:
$|p(t)\rangle=(p_1(t),...p_i(t)...p_N(t))$, so that $\langle i|p(t)\rangle= p_i(t)$
The probabilities $p_i(t)$ obey the following equations \cite{redner-book,gardiner-book,bezrukov-sites-2005}
\begin{eqnarray}\label{kinetics_sp}
\frac{d}{dt}{p}_i(t)=r(p_{i-1}+p_{i+1}-2p_i)\;\;\text{for}\;\;1<i<N
\end{eqnarray}
with the boundary conditions
\begin{equation}\label{kinetics_sp_bc}
\frac{d}{dt}{p}_1(t)=-(r_o+r)p_1+rp_2)\ \ \ \ \text{and} \ \ \ \ \frac{d}{dt}{p}_N(t)=-(r_o+r)p_N+rp_{N-1}.
\end{equation}
Equations~(\ref{kinetics_sp},\ref{kinetics_sp_bc}) can be written in a matrix form as
\begin{equation}\label{equation-for-n-1-sp--prl}
\frac{d}{dt}|p(t)\rangle=\hat{M}\cdot |p(t)\rangle,
\end{equation}
with
\begin{equation}
M_{i,i}=-2r \ \ \ \ \text{and} \ \ \ \ M_{i,i\pm 1}=r \ \ \ \ \text{for} \ \ \ \ 1<i<N,
\end{equation}
and
\begin{equation}
M_{1,1}=-r-r_o; \ \ \ \ M_{N,N}=-r-r_o; \ \ \ \ M_{1,2}=r; \ \ \ \ M_{N,N-1}=r.
\end{equation}
\subsubsection{Explicit solution of single particle equations in terms of matrix elements}
Here we re-derive the solutions obtained in the main text, using the standard methods of linear algebra \cite{arfken-book}. Assume an arbitrary  Markov process that can be in $N$ states (such as defined by equation (3)). Time evolution of its probability distribution $\vec{p}(t)=(p_1,p_2,...,p_N) $ and can be described by the following matrix equation (the equation (3) is an example):
\begin{equation}
\dot{\vec{p}}=\hat{U}\cdot\vec{p}
\end{equation}
where $\vec{p} $ is an $N$-dimensional vector, and $\hat{U} $ is an $N\times N $ matrix.
Let us denote the eigenvalues of the matrix $\hat{U} $ as $\omega_1...\omega_i...\omega_N $ and the corresponding eigenvectors as $\vec{v}^1,..., \vec{v}^i,...\vec{v}^N $. Then the general solution is
\begin{equation}
\vec{p}(t)=\sum_{j=1}^{N}a_j\vec{v}^je^{\omega_j t},
\end{equation}
where $a_1...a_N$ is a set of numerical coefficients. In other words,
\begin{equation}
p_i(t)=\sum_{j=1}^{N}a_jv^j_ie^{\omega_j t}.
\end{equation}
The coefficients $a_j $ can be determined from the initial condition:
\begin{equation}
p_i(0)=\sum_{j=1}^{N}a_jv^j_i,
\end{equation}
which can be written as
\begin{equation}
\vec{p}(0)=\hat{V}\cdot\vec{a},
\end{equation}
where $$\hat{V}=\left(\begin{array}{c}
                  \vec{v}^1 \\
                  ... \\
                  \vec{v}^i \\
                  ... \\
                  \vec{v}^N
                \end{array}\right)^T
$$ so that $V_{ij}=v^j_i $.
Finally
\begin{equation}
\vec{a}=\hat{V}^{-1}\cdot\vec{p}(0)
\end{equation}
and
\begin{equation}
a_k=\sum_{j=1}^{N}(V^{-1})_{kj}p_j(0)
\end{equation}

For the initial condition $p_i(0)=\delta_{i,1} $, we get $a_k=(V^{-1})_{k1}.$
Now, the matric $U$ is diagonalized to its diagonal form

$$\hat{W}=\left(\begin{array}{ccc}
                  w_1 & 0 & ...\;\;\;0 \\
                  .....&.....&....\\
                 0\;\; ... & w_i & ...\;\;\;0\\
                 .....&....&....\\
                  0\;\; ... & 0 & w_N \\
                \end{array}\right)
$$

by the transformation  $W=V^{-1}UV $, or equivalently $U=VWV^{-1} $ \cite{arfken-book}.

Thus, the probability to be in state $i$ at time $t$ is
\begin{eqnarray}
p_i(t)=\sum_{j=1}^{N}a_jv^j_i\exp{w_jt}=\sum_{j=1}^{N}(V^{-1})_{j1}V_{ij}e^{w_jt}=\left(Ve^{Wt}V^{-1}\right)_{i1}=\left(e^{Ut}\right)_{i1}
\end{eqnarray}
and in particular,
\begin{eqnarray}
p_N(t)=\left(e^{Ut}\right)_{N1}
\end{eqnarray}

The probability flow to exit to the right is $r_op_N(t)$, and the total probability of exit to the right is

\begin{align}
P_{\rightarrow}&=r_o\int_0^{\infty}p_N(t)dt=-r_o\sum_{j=1}^{N}a_jv^j_N\frac{1}{w_j} \\ \nonumber
&=-r_o\sum_{j=1}^{N}(V^{-1})_{j1}V_{Nj}\frac{1}{w_j}
\left(VW^{-1}V^{-1}\right)_{N1}=-r_oU^{-1}_{N1}
\end{align}

in agreement with equation (2) in the main text. This result can be also obtained using the following reasoning. Instead of considering a single particle hopping through the states, starting at the position $1$, let us consider the steady state where a flux $J$ enters to a position $1$, with a steady state probability distribution $\vec{p} $. Then the probability to exit to the right is the ratio of the transmitted flux to the entrance flux: $r_op_N/J $. We have
\begin{equation}
0=\hat{U}\cdot \vec{p}+\vec{J}
\end{equation}
and therefore $\vec{p}=-U^{-1}\cdot\vec{J} $ so that $p_N=-U^{-1}_{N1}$ because $J_i=J\delta_{i,1} $.

The probability distribution of exit times to the right is simply $r_op_N(t)$ \cite{bezrukov-sites-2005} and any moment of it can be calculated easily.
For instance, the mean first passage time to exit to the right is:
\begin{equation}
\bar{T}_{\rightarrow} = r_o\int_0^{\infty}tP_N(t)dt=r_o\sum_{j=1}^{N}(V^{-1})_{j1}V_{Nj}\frac{1}{w_j^2}=r_o\left(V(W^2)^{-1}V^{-1}\right)_{N1}=r_o\left(U^2\right)^{-1}_{N1}
\end{equation}
in agreement with the equation (3) in the main text.

\subsection{Steady state}
In the case of current  $J$ impinging on the channel entrance, one can describe the system in terms of average site occupancies $n_i$, whose kinetics is described in the mean field approximation by the following equations  \cite{schutz-review-2005,chou-PRL-single-file-1998,zilman-BJ}.
\begin{align}\label{kinetics_with_exclusion-prl}
\frac{d}{dt}{n}_i &=rn_{i-1}(1-\frac{n_i}{m})+rn_{i+1}(1-\frac{n_i}{m})-rn_i(1-\frac{n_{i-1}}{m})-rn_i(1-\frac{n_{i+1}}{m})\\ \nonumber
&=r(n_{i-1}+n_{i+1}-2n_i) \ \ \ \ \text{for} \ \ \ \ 1<i<N.
\end{align}
where $m$ is the maximal site occupancy. The boundary conditions at sites $1$ and $N$ are
\begin{eqnarray}\label{kinetics_with_exclusion-2}
&&\frac{d}{dt}n_1=-(r+r_o)n_1+rn_2+J(1-\frac{n_1}{m}) \nonumber \\
&&\frac{d}{dt}n_{N}=-(r+r_o)n_{N}+rn_{N-1}.
\end{eqnarray}
In a matrix form:
\begin{eqnarray}\label{equation-for-n-1-prl}
\frac{d}{dt}|n(t)\rangle=\hat{M}^J\cdot |n(t)\rangle+\vec{J}
\end{eqnarray}
where the matrix $\hat{M}^J$ is the same as $\hat{M}$ with the only change $\hat{M}^J_{1,1}=-J/m-r-r_o$ and $\vec{J}=(J,0,...0)$.
Note that for an internally uniform channel (as the one described in Fig. 1) the mean-field equations (Eqs.~(\ref{kinetics_with_exclusion-prl},\ref{equation-for-n-1-prl})) are exact \cite{schutz-review-2005,chou-PRL-single-file-1998}.

The steady state density profile can be obtained from Eq.~(\ref{equation-for-n-1-prl}) as $|n\rangle^{ss}=-\left(\hat{M}^J\right)^{-1}\cdot \vec{J}$, or more specifically as:
\begin{eqnarray}\label{equation-ni}
n^{ss}_i=\frac{J\left(1+(N-i)\frac{r_o}{r}\right)}{r_o\left(2+(N-1)\frac{r_o}{r}\right)+\frac{J}{m}\left(1+(N-1)\frac{r_o}{r}\right)}.
\end{eqnarray}
The average exit flux to the right is $J_{\rightarrow}=r_on^{ss}_N$. This together with Eq.~(\ref{equation-ni}) yield the probability of an individual particle within the flux to exit to the right:
\begin{equation}\label{p-exit-ss-prl-1}
P^{ss}_{\rightarrow}=\frac{J_{\rightarrow}}{J(1-n_1/m)}=\frac{1}{2+(N-1)r_o/r}.
\end{equation}
As already established before, the exit probability of individual particles to exit to the right is the same as in the single-particle case (at least for uniform channels), even though they are interfering with each other's passage through the channel \cite{zilman-BJ}.

However, crowding does influence transport and is manifested in obstruction of the entrance site. The transport \textit{efficiency}, defined as the ratio of the exit flux to the right $J_{\rightarrow}$ to the total impinging flux $J$, $\text{Eff}_{\rightarrow}=\frac{J_{\rightarrow}}{J}$,
decreases with $J$ due to jamming at the entrance.

\begin{equation}\label{p-exit-ss-J-prl}
\text{Eff}_{\rightarrow}=\frac{J_{\rightarrow}}{J}=\frac{r_o}{2r+(N-1)r_o+J(1+(N-1)r_o/r)/m}.
\end{equation}

\subsection{Derivation of the analytical expressions for the mean exit times in the jammed regime}

The boundary conditions of equation (8) of the main text, describing the probability of the tagged particle are:
\begin{eqnarray}\label{kinetics_with_exclusion-2-ss-prl}
&&\frac{d}{dt}{p}_1=-r_op_1-rp_1(1-n^{ss}_2/m)+rp_2(1-n^{ss}_1/m)\nonumber\\
&&\frac{d}{dt}{p}_{N}=-r_op_{N}-rp_{N}(1-n^{ss}_{N-1}/m)+rp_{N-1}(1-n^{ss}_{N}/m)
\end{eqnarray}
Using the matrix form of Eq.~(9) of the paper, the elements of the matrix $M^{ss}$ are given by
\begin{align}
M^{ss}_{i,i}&=-r(2-n^{ss}_{i-1}/m-n^{ss}_{i+1}/m)=-2r(1-n^{ss}_i/m);\\ \nonumber 
\text{and} \ \ \ \ M^{ss}_{i,i\pm 1}&=r(1-n^{ss}_i/m) \ \ \ \ \text{for} \ \ \ \ 1<i<N,
\end{align}
and
\begin{align}
M^{ss}_{1,1}=-r(1-\frac{n^{ss}_1}{m})-r_o; \ \ M^{ss}_{N,N}=-r(1-\frac{n^{ss}_{N-1}}{m})-r_o; \\ \nonumber
M^{ss}_{1,2}=r(1-\frac{n^{ss}_1}{m}); \ \ M^{ss}_{N,N-1}=r(1-\frac{n^{ss}_N}{m}).
\end{align}
The average (over particles actually exited to the left) time to exit to the left is
\begin{equation}
\overline{T}^{ss}_{\leftarrow}=r_o\left\langle 1\right\vert \left(
\left(  M^{ss}\right)  ^{-1}\right)  ^{2}\left\vert 1\right\rangle/P_{\leftarrow},
\end{equation}
where
\begin{equation}
P_{\leftarrow}=1-P_{\rightarrow}=\frac{1+\left(  N-1\right)  r_o/r}{2+\left(  N-1\right)
r_o/r}.
\end{equation}
In order to obtain an explicit expression for $\overline{T}^{ss}_{\leftarrow}$ we define
\begin{equation}
|W\rangle=D  \left(  M^{ss}\right)  ^{-1}\left\vert
1\right\rangle
\end{equation}
and
\begin{equation}
\langle Q|=D  \left\langle 1\right\vert \left(  M^{ss}\right)
^{-1}.
\end{equation}
In the equations above we introduced the notation for the determinant of $M^{ss}$, $D\equiv det\left(M^{ss}\right)$.
The elements of these vectors are given by
\begin{equation}
W_{n}=r^{N-2}\left(  A_{N}r+nB_{N}r+\left(  N-n\right)  r_o\right)
{\displaystyle\prod\limits_{k=2}^{N-1}}
\left(  A_{N}+kB_{N}\right)
\end{equation}
and
\begin{equation}
Q_{n}=r^{N-2}\left(  A_{N}r+nB_{N}r+\left(  N-n\right)
r_o\right)  \frac{%
{\displaystyle\prod\limits_{k=1}^{N-1}}
\left(  A_{N}+kB_{N}\right)  }{A_{N}+nB_{N}}.
\end{equation}
In the above expressions
\begin{equation}
A_{N}=\frac{r_o\left(  2r+\left(  N-1\right)  r_o-J/m\right)  }{(J/m)\left(
r+\left(  N-1\right)  r_o\right)  +r_o\left(  2r+\left(  N-1\right)
r_o\right)  }%
\end{equation}
and
\begin{equation}
B_{N}=\frac{Jr_o/m}{(J/m)\left(  r+\left(  N-1\right)  r_o\right)  +r_o\left(
2r+\left(  N-1\right)  r_o\right)  }.
\end{equation}
The mean escape time to the left is then
\begin{align}
&\overline{T}^{ss}_{\leftarrow}P_{\leftarrow}=\frac{r_{o}}{D^{2}}%
{\displaystyle\sum\limits_{n=1}^{N}}
W_{n}Q_{n}\\ \nonumber
&=\frac{r_o}{D^{2}}%
{\displaystyle\sum\limits_{n=1}^{N}}
r^{2N-4}\left(  A_{N}r+nB_{N}r+\left(  N-n\right)  r_o\right)  ^{2}%
\frac{\left(  A_{N}+B_{N}\right)  }{A_{N}+nB_{N}}%
{\displaystyle\prod\limits_{k=2}^{N-1}}
\left(  A_{N}+kB_{N}\right)^{2}.
\end{align}
Using the notations above we can express $D$ as:
\begin{equation}
D=r^{N-2}r_o\left(  2A_Nr+\left(  N+1\right)
B_Nr+\left(  N-1\right)  r_o\right)
{\displaystyle\prod\limits_{k=2}^{N-1}}
\left(  A_{N}+kB_{N}\right).
\end{equation}
Substituting $D$ into the expression for $\overline{T}^{ss}_{\leftarrow}$ one obtains%
\begin{equation}
\overline{T}^{ss}_{\leftarrow}P_{\leftarrow}=\frac{\left(  A_{N}+B_{N}\right)
{\displaystyle\sum\limits_{n=1}^{N}}
\frac{\left(  A_{N}r+nB_{N}r+\left(  N-n\right)  r_o\right)  ^{2}}%
{A_{N}+nB_{N}}}{r_o\left(  2A_{N}r+\left(  N+1\right)  B_{N}r+\left(
N-1\right)  r_o\right)  ^{2}}.
\end{equation}
Performing the summation, we get for the average time to exit the channel to the left%
\begin{align}
\frac{\overline{T}^{ss}_{\leftarrow}P_{\leftarrow}}{\left(  A_{N}+B_{N}\right)}&=\frac{
\left(N(B_N r-r_o) (2 A_N (B_N r+r_o)+B_N (B_N (N+1) r+(3 N-1)r_o))\right)}{2 B_N^2r_o\left(  2A_{N}r+\left(  N+1\right)  B_{N}r+\left(
N-1\right)  r_o\right)  ^{2}}\nonumber \\
&+\frac{
r_o (A_N+NB_N)^2\left(\psi\left(\frac{A_N}{B_N}+N+1\right)-\psi\left(\frac{A_N+B_N}{B_N}\right)\right)}{B_N^3\left(2A_{N}r+\left( N+1\right) B_{N}r+\left(
N-1\right)  r_o\right)  ^{2}}.
\end{align}
Here $\psi\left(  x\right)  =\frac{d\ln\left(  \Gamma\left(  z\right)
\right)  }{dz}|_{z=x}$, where $\Gamma (x)$ is the $\gamma$-function.

Substituting the expressions for $A_N$ and $B_N$, we get the explicit expression for the
average time as
\begin{align}
&\overline{T}^{ss}_{\leftarrow}   =-\frac{N\left(2r_o\left(2r+r_o\left(N-1\right)\right)+\frac{J}{m}\left(4r+3r_o\left(N-1\right)\right)\right)}{2\left(J/m\right)^2\left(r+r_o\left(N-1\right)\right)} \nonumber \\
& +\frac{(J/m)\left(r+r_o\left(N-1\right)\right)+r_o\left(2r+r_o\left(N-1\right)\right)}{(J/m)^3 r_o \left(r+r_o\left(N-1\right))\right)}\left(\psi \left(N+\frac{2 r+\left(N-1\right)
   r_o}{J/m}\right)-\psi \left(\frac{2 r+(N-1) r_o}{J/m}\right)\right).
\end{align}
In the main text we use the definitions of the probabilities to exit to the right/left in order to simplify this cumbersome expression. In the single particle limit, $J\to 0$, the expression for $\overline{T}^{ss}_{\leftarrow}$ reduces to the previously obtained single particle expression (Eq.~(6) in the paper).
At the other extreme when the input flux $J\to \infty$, the mean escape time is
\begin{equation}
\lim_{J \to \infty}\overline{T}^{ss}_{\leftarrow}=\frac{r+(N-1)r_o}{r_o(2r+(N-1)r_o)}=\frac{P_{\leftarrow}}{r_o}.
\end{equation}
The mean time to exit to the right and the mean trapping time can be obtained in a similar fashion.

%\bibliographystyle{apsrev}
%\bibliography{transport}

\end{document}